# Investigation of thrust augmentation and nozzle exit pressure in starting jets from a tube nozzle

Ali A. Moslemi[1]

Pulsed Flow Solutions, The Woodlands, TX

**Abstract:** It has been widely accepted that nozzle exit over-pressure is responsible for thrust augmentation in starting jets over steady jets and a few pulsed jet propulsion systems have been developed based on this finding. However, no other study has been conducted to confirm the nozzle exit over-pressure effect. In this paper, thrust and nozzle exit pressure are numerically investigated under different jet velocity programs (negative slope (NS), positive slope (PS) and impulsive). Starting jets are generated for fluid slug length to diameter (*L/D*) ratios of *2-5* at each jet velocity program. In contradiction to the findings presented in [P. Krueger and M. Gharib, J. *Phys. Fluids,* 15, 1271 (2003)] the results in this paper reveal that the time-averaged thrust due to nozzle exit pressure in starting jets from a tube nozzle is negligible or negative resulting in no thrust augmentation when compared to steady jets with hypothetical identical velocity programs. Also, nozzle exit experiences both over-pressure and under-pressure in accelerating and decelerating portions of jet velocity programs respectively in both NS and PS velocity programs. Furthermore, it is found that secondary counter-rotating vortex is generated during decelerating portion of jet velocity programs which may be linked to the nozzle exit under-pressure explaining the reduced thrust during starting jets generated by different velocity programs.

## 1. Introduction

Unsteady jets including starting and pulsed jets have been extensively studied for many reasons including understanding the locomotion of aquatic creatures like jellyfish and squid as well as thrust augmentation of pulsed jets over steady jets (Weihs 1977, Kruger & Gharib 2003, 2005 and Athanassiadis & Hart 2016). Also, few bioinspired propulsion systems have been designed based on pulsed jets as reported in Moslemi & Krueger 2010, Ruiz et al. 2010 and Serchi et al. 2013. A pulsed jet consists of a series of its fundamental unit called starting jet, with a finite duration of no flow between them. Therefore, studying the hydrodynamic properties of starting jets including thrust and impulse provides a foundation for understanding pulsed jets and their applications. Due to their similarity and close relationship, the thrust and nozzle exit pressure for both starting and pulsed jets are considered in the following literature review.

Weihs 1977 raised the question of propulsive benefit of pulsed jets over steady jets motivated by possible evolutionary incentive of this swimming technique used by aquatics creatures like Salps and jelly fish. He derived an analytical equation to calculate the thrust augmentation of pulsed jet over steady jet with equal mass flow rates. In the provided analysis, it is shown that this ration is always greater than unity and can reach 1.25 to 1.6 with some assumptions including that the ration of added mass to vortex ring mass is about one. Vortex ring is a key flow feature which is generated during starting and pulsed jets.

Sarohia et al., 1981 studied the thrust augmentation of pulsed jets in conjunction with ejectors. In their experiments, enough information is not provided about how pulsed jets are created but it is inferred that a velocity oscillation is imposed on a steady jet considering the presented pulsed jet generator (see figure 1 of the reference) which works based on momentarily restricting the air flow. In their experiments, some limited data of thrust augmentation of pulsed jets without presence of ejector is provided. Their tests showed that the thrust augmentation was found to be independent of the pulsing frequency but was proportional to the pulsing jet velocity fluctuation which is the ratio of amplitude of jet velocity variation to the time-averaged jet velocity. Maximum thrust augmentation reached about 4% which was obtained at maximum jet velocity fluctuation of 17% indicating the dependence of thrust augmentation on pulsation amplitude fluctuation.

Anderson and Demont 2000 conducted high-speed, high-resolution digital recordings of squid swimming to determine accurate swimming kinematics, body deformation and mantle cavity volume. Employing these data in an unsteady Bernoulli equation, they calculated intramantle (mantle cavity) pressure of swimming squid during the jet ejection. It was found that both positive and negative (gage) pressures exist in squid intramantle during accelerating and decelerating portions of the jet respectively. However, it was not specified if the time-averaged of this pressure is positive or negative during the whole jet ejection. It is worth noting that while the pressure in squid intramantle and the pressure at its jet orifice (which is the focus of this study) are different, but they are closely related. Recently, Krieg and Mohseni 2015 presented an analytical equation relating the pressures in cavity and at orifice exit of deformable and axisymmetric jet producing cavity bodies but did not provide any numerical data for pressures.

---

[1] Email address for correspondence: amoslemi@gmail.com

Motivated by lack of enough data for pulsed jet augmentation over steady jet, Krueger and Gharib 2003 and 2005 conducted a series of experiments to study flow dynamic properties including thrust and impulse for both starting and fully pulsed jets using DPIV (Digital Particle Image Velocimetry) and direct force measurements. They measured thrust augmentation under different triangular velocity programs (variation of average jet velocity during pulse time) and different stroke-to-diameter ratio ($L/D$) which is the length of fluid slug ejected during the pulse to the nozzle diameter. Two measurement metrics were defined to compare the generated thrust from pulsed jets to steady jets. The difference between these two methods relates to how to calculate the thrust from equivalent steady state jet. The first metric is based on comparison of thrust generated from pulsed jet with that of steady jet with equal time-averaged mass flux similar to what was used in Weihs 1977 and Sarohia et al. 1981. The second metric calculates the steady jet thrust based on velocity profile formed at the nozzle exit under the same velocity program but without any unsteady effect. The latter ratio is dubbed the intermittent-jet normalization in Krueger & Gharib 2005 and it is used in this study for thrust augmentation comparison. The results of starting jet tests showed that thrust augmentation was higher than one in all tested parameters based on both metrics and revealed a relative maximum near the $L/D= 4$ in which vortex pinch-off occurs based on previous tests performed by Gharib et al. 1998. The thrust augmentation ratio (intermittent-jet normalization ratio) reached about 1.7 at $L/D= 2$ and then it decreased for $L/D>4$ signifying the effect of vortex ring pinch-off. Krueger & Gharib 2003 attributed the improved thrust of starting jets over steady jets to the nozzle exit over-pressure due to added and entrained masses during vortex ring formation. The pressure impulse contribution to the total impulse was found to be as much as 42% of the total impulse for cases involving isolated vortex rings ($L/D<4$). However, it was acknowledged that the detailed transient measurements of nozzle exit pressure and the respected pressure impulse were necessary to verify the findings. This is a significant shortcoming and is one of focuses of this study.

For the tests related to fully pulsed jets, similar trend was observed with 90% thrust augmentation (intermittent-jet normalization ratio) at $L/D$=2 and pulsing duty cycle ($Sr_L$) of 0.15. Also, this augmentation decreased as $L/D$ (similar to starting jet tests) and $Sr_L$ increased. $Sr_L$ varies between 0 and 1 for fully pulsed jet and determines the separation between pulses. As it increases, the vorticity from preceding pulses becomes closer to the nozzle exit which requires less fluid to be accelerated by the issuing pulse and reduces nozzle exit over-pressure (Krueger & Gharib 2005).

Choutapalli et al. 2009 studied the thrust augmentation of pulsed air flow in conjunction with ejectors using DPIV and direct thrust measurements. Also, direct thrust measurements were performed to find pulsed jet thrust augmentation over steady jet with equivalent average mass flux. A pitot total probe was placed at the center of the nozzle exit to monitor the total pressure of the jet during pulsation. The results showed that the thrust ratio based on equal mass flux rate increased from 1.08 to 1.15 when $Sr_L$ increased from 0.03 to 0.11 showing direct dependence of thrust augmentation to $Sr_L$. The thrust augmentation from nozzle exit over-pressure was reported as 7% of the total thrust at $Sr_L = 0.106$. The reported augmentation values are similar to those reported in Sarohia et al., 1981 but they are much lower than the augmentation values reported in Krueger & Gharib 2005. To justify the low augmentation values, Choutapalli 2007 mentioned that in his tests, unlike the tests described in Krueger & Gharib 2005, the generated pulsed jets did not reach zero velocity between pulses due to the leak in the experimental setup. As a result, the flow was not stationary in the beginning of each pulse causing less thrust contribution from nozzle exit over-pressure. Also, it is mentioned that the Reynolds number in the conducted tests is an order of magnitude higher than that of Krueger & Gharib 2003, 2005 and that can be another factor why the augmentation values are different, but no further explanation is provided. It is worth noting that Choutapalli et al., 2009 reported that the *L/D* was larger than 9.4 in all of the tested cases and acknowledged that the dynamics of the pulsed jets in their study would be different from those available in the previous studies. Despite the reliance of Choutapalli 2007 to the effect of nozzle exit over-pressure to justify thrust augmentation, no (static) pressure measurement was provided to support such a claim and only total pressure measurements at the nozzle exit during pulsation have been reported. Presented total pressure measurements showed oscillating behavior of pressure during jet pulsation. It is difficult to extract the static pressure variation from this data without having the information about the test conditions. However, Choutapalli et al. 2009 stated that nozzle exit over-pressure occurs only when the jet velocity increases with time but did not comment about it when the jet velocity decreases during pulsation.

In pursuit of finding an analytical model to determine nozzle exit pressure during pulsation, Krueger 2001 derived an equation for the nozzle exit pressure and total pressure impulse using unsteady Bernoulli and momentum equations. With a similar approach, Krieg & Mohseni 2013 derived a similar equation that estimates the nozzle exit pressure based on the knowledge of the velocity field at the nozzle exit plane. Based on the mentioned methods, another equation is presented here to shed some light on the nozzle exit pressure change during a starting jet. Figure 1 shows a piston-cylinder mechanism similar to what was used for generating starting jets in Krueger & Gharib 2003.

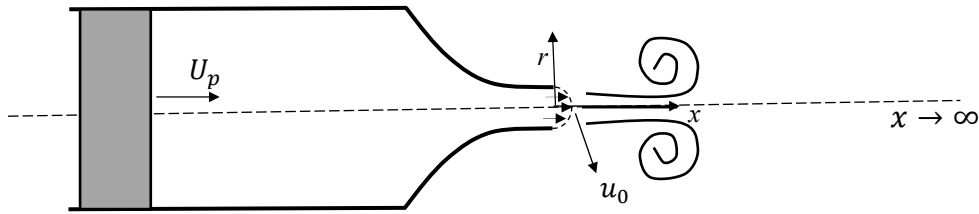

Figure 1. The piston-cylinder mechanism for generating a starting jet

Figure 1 shows an axisymmetric starting jet exiting from the nozzle. Following Krueger 2005 and Krieg & Mohseni 2013, the flow at the centerline is considered irrotational and therefore unsteady Bernoulli equation is employed. The radial velocity at the nozzle exit centerline is zero considering the axisymmetric condition of the flow. Applying unsteady Bernoulli's equation at the nozzle exit centerline and far-field boundaries gives:

$$\frac{\partial \emptyset_0}{\partial t} + \frac{1}{2}u_0^2 + \frac{p_0}{\rho} = \frac{\partial \emptyset_\infty}{\partial t} + \frac{p_\infty}{\rho} \tag{1}$$

Where $\phi$ is the velocity potential and also, $p_0$ and $p_\infty$ are respectively pressures at the nozzle exit center and far field which is equal to ambient pressure. $U_p$ and $u_o$ refer to piston speed and jet velocity at nozzle exit center respectively. The velocity potential can be determined from the axial velocity by definition of the potential function, that is $u = \frac{\partial \emptyset}{\partial x}$, since the axis of symmetry is irrotational. Velocity potential at the nozzle exit center, $\emptyset_0$, and velocity potential at the far filed, $\emptyset_\infty$, can be obtained by integrating the flow axial velocity along the centerline as following:

$$\emptyset_0 = \emptyset_\infty - \int_0^\infty u dx \tag{2}$$

Taking time derivative of (2) and inserting it in (1) yields:

$$\frac{p_0 - p_\infty}{\rho} = \int_0^\infty \frac{\partial u}{\partial t} dx - \frac{1}{2}u_0^2 \tag{3}$$

Equation (3) provides a relatively simple formulation for nozzle exit over-pressure (at center point) based on the integration of rate of the flow velocity change with respect to time along the centerline as well as time-dependent nozzle exit velocity. It is difficult to determine how exactly nozzle exit pressure changes without having the knowledge of flow field velocity during starting jet. However, it is possible to gain some insights from this equation considering the type of jet velocity program employed. It is important to note that the first term of equation (3) on the right, $\int_0^\infty \frac{\partial u}{\partial t} dx$, is influenced by both the jet velocity and the velocity induced by the formation of the primary vortex ring. But, for simplicity in the analysis, it is assumed that the jet velocity has the dominant effect in the integral. Assuming a triangular jet velocity program similar to those used in Krueger & Ghairb 2003 and 2005, it is reasonable to assume that the first term at right side of equation (3), $\int_0^\infty \frac{\partial u}{\partial t} dx$, is mainly positive when the jet is in accelerating mode since $\frac{\partial u}{\partial t}$ is positive at each point along the centerline considering that the ambient flow is stagnant before jet ejection. If $\int_0^\infty \frac{\partial u}{\partial t} dx > \frac{1}{2} u_0^2$, then nozzle exit gauge pressure would become positive which translates to the nozzle exit over-pressure at the respective time of jet ejection. This behavior has been reported by Choutapalli et al. 2009 without providing any experimental data as stated before. Similarly, $\int_0^\infty \frac{\partial u}{\partial t} dx$ should become negative as the jet decelerates resulting in negative nozzle exit gauge pressure or under-pressure since there are two negative terms at the right side of equation (3). Therefore, nozzle exit is expected to experience positive-negative gauge pressures similar to a sinusoidal wave during starting jet based on the presented assumptions. If this holds true, it will be in contradiction to what Krueger & Ghairb 2003 and 2005 concluded emphasizing only nozzle exit over-pressure portion. Therefore, it is essential to obtain the nozzle exit pressure field during pulsation to verify the presented model.

Krieg & Mohseni 2013 calculated the impulse and nozzle exit pressure using DPIV and a newly analytical formulation (as mentioned) for starting jets ejected from both a tube and orifice for mainly two $L/D$s equal to *2.4* and *6.9*. The results related to the starting jets ejected from a tube using an impulsive jet velocity program at $L/D$= *2.4* revealed some contradiction with the results of Krueger & Ghairb 2003. The nozzle exit over-pressure was only present at the initial moments of the jet when the jet velocity program was accelerating in agreement to the presented model for the nozzle exit pressure. The nozzle exit (gauge) pressure becomes almost zero as the jet velocity reaches its final value and remains so until the end of jet termination. It is worth noting that the test results for the orifice nozzle were quite different since the nozzle exit over-pressure was always present during starting jets. This was attributed to the presence of radial velocity during vortex ring formation in the orifice nozzle whereas it was very weak in the tube nozzle. Regarding this effect, more details and discussion are provided in Krieg & Mohseni 2013 which are not repeated here. The focus of this paper is given to the starting jets from a tube nozzle.

Recently, Zhang et al. 2020 computed the thrust from laminar starting jets ejected from a tube nozzle using CFD and showed that the thrust due to nozzle exit over-pressure is negligible compared to the momentum thrust for impulsive jet velocity programs. Also, they calculated the total thrust for different triangular jet velocity programs but did not report the thrust contribution from nozzle exit pressure since the focus of study was propulsion efficiency of starting jets.

Considering the presented literature review signifying the lack of data for nozzle exit pressure during starting jets (from a tube nozzle), it is important to investigate this parameter in a systematic manner to confirm its effect on thrust augmentation of starting jets. Computational Fluid Dynamics (CFD) is used to create starting jets similar to those generated in Krueger & Ghairb 2003 and then the pressure at the nozzle exit is computed for different triangular and impulsive jet velocity programs. Also, thrust is calculated for the respective cases and is compared to the thrust of its equivalent steady jet based on identical velocity profiles at the nozzle exit without any unsteady effect. Finally, the contribution of pressure thrust to the total thrust is obtained and discussed for each jet velocity program.

The remainder of the paper is organized as follows. In Section 2, the definitions for numerical test parameters including jet velocity programs and thrust calculation are provided. Section 3 introduces the computational set-up and mesh independence study. Section 4 presents the CFD modeling results including vortex ring formation, nozzle exit pressure and thrust augmentation results under different velocity programs. Section 5 summaries and concludes the paper with final remarks.

## 2. Numerical test parameters

Stroke length-to-diameter ratio (*L/D*) is a critical parameter for vortex ring formation generated during starting jets. It is defined as the ratio of the total ejected fluid slug to the nozzle exit diameter, namely

$$\frac{L}{D} = \frac{1}{D} \int_0^T U_j \, dt \tag{4}$$

Where $U_j$ is the jet velocity averaged over the nozzle cross-section and *T* is the jet duration. Stroke ratios of *L/D= 2,3,4,* and *5* were simulated using three different velocity programs. Figure 2 shows the velocity programs including negative slope, positive slope and impulsive velocity programs.

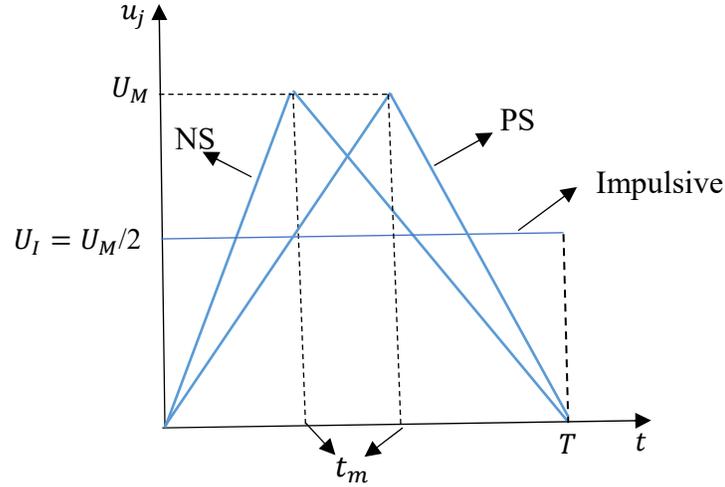

Figure 2. Triangular negative slope (NS), positive slope (PS) and impulsive velocity programs for a starting jet

Following Krueger & Gharib 2005, a triangular jet velocity program is defined as negative slope (NS) or positive slope (PS) if $\frac{t_m}{T}$ is smaller or greater than 0.5 respectively. $t_m$ refers to the time that jet reaches its maximum velocity. The jets in the NS and PS programs have mainly decelerating and accelerating velocities respectively. The impulsive velocity program has the magnitude of $U_M/2$ to enforce the identical mass flux rate of other two velocity program since the time-averaged mass flux of starting jet is proportional to the surface areas under the triangles shown in figure 2 which is $\frac{U_M T}{2}$.

The time-averaged thrust in a starting jet, $F_{sj}$, is the summation of time-averaged thrust due to the jet momentum and nozzle exit pressure expressed as

$$F_{sj} = F_u + F_p = \frac{\rho}{T} \int_0^T \int_A u_j^2(t) dS dt + \frac{1}{T} \int_0^T \int_A (p_0 - p_\infty) dS dt \tag{5}$$

To determine the thrust augmentation, the same approach suggested by Krueger & Gharib 2005 is employed. Dubbed the intermittent-jet normalization, in this method, the starting jet thrust is compared with the thrust of a hypothetical jet of identical velocity program but without considering the jet transient effects like nozzle exit over-pressure. In other words, the thrust of the hypothetical jet is simply equal to $F_u$ resulting in the thrust augmentation ratio as

$$F_A = \frac{F_{sj}}{F_u} = 1 + \frac{F_p}{F_u} \tag{6}$$

The above ratio is calculated for different *L/D* and velocity programs. Using this method, the effect of nozzle exit over-pressure in thrust augmentation is fully captured.

## 3. CFD modeling setup and validation

The experimental setup provided in Krueger & Gharib 2003 is adopted for numerical modeling of starting jets to compute the thrust augmentation ratio and obtain the nozzle exit over-pressure. Figure 3 illustrates a half cross-section view of the 3D model of the studied nozzle with generated computational mesh. Quadrilateral elements in an unstructured mesh with different resolutions are used to increase the accuracy of numerical analysis for the regions with high gradient of flow quantities. Cut cell elements are used near the curved solid boundaries to retain a boundary conforming grid. Following Krueger & Gharib 2003, the minimum distance of nozzle from any boundaries is *25D* where *D=0.5 in. (1.27 cm)* is the nozzle diameter. A fine mesh is used near the nozzle exit area to capture all details of vortex ring formation. This area is extended up to *2D* radially and *8D* axially considering the range of *L/D=2-5* studied in this paper. The piston motion is modeled as a uniform velocity inlet following Rosenfeld, et al. 2009. Applying symmetric boundary conditions, only a quarter of this 3D geometry is modeled to save computational time. It is worth noting that the used CFD package was designed for 3D models and employing a 2D axisymmetric model was not possible in this software.

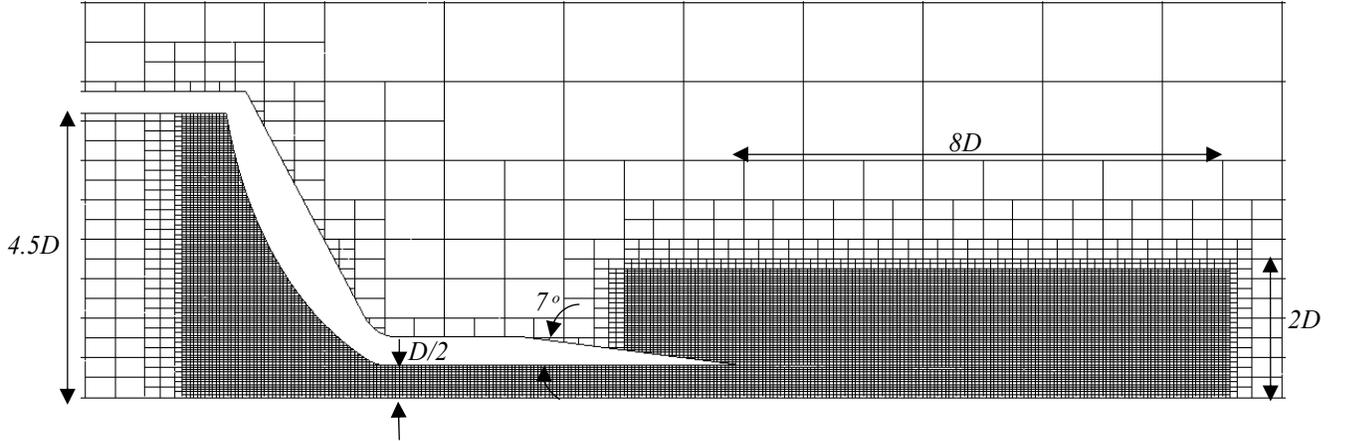

Figure 3. Half cross-section view of the 3D model of the studied nozzle with generated computational mesh

The Reynolds number based on the $U_M$ and nozzle diameter is *1.3×10⁴* indicating that the flow transitions from laminar to turbulent as it moves from cylinder to nozzle section (because of different flow velocities) during jet ejection in the numerical tests. Therefore, a laminar/turbulent model is used to account for this transition using one system of equations. To model the turbulent flow, the Favre-averaged Navier-Stokes equations are used. The *k-ε* model is used to close this system of equations and a commercial CFD package is employed to solve these equations using the cell-centered finite volume method. The conservation form of mass and momentum equations in the cartesian coordinate system can be written as follows:

$$\frac{\partial \rho}{\partial t} + \frac{\partial}{\partial x_i}(\rho u_i) = 0$$

$$\frac{\partial \rho u_i}{\partial t} + \frac{\partial}{\partial x_j}(\rho u_i u_j) + \frac{\partial p}{\partial x_i} = \frac{\partial}{\partial x_j}(\tau_{ij} + \tau_{ij}^R) - \rho g_i \quad i = 1,2,3 \tag{7}$$

where *u* and *ρ* refer to the fluid velocity and density respectively and $g_i$ is the gravitational acceleration component along the *i-th* coordinate axis. Also, $\tau_{ij}$, the viscous shear stress tensor is defined as:

$$\tau_{ij} = \mu \left( \frac{\partial u_i}{\partial x_j} + \frac{\partial u_j}{\partial x_i} - \frac{2}{3}\delta_{ij}\frac{\partial u_k}{\partial x_k} \right) \tag{8}$$

Following Boussinesq assumption, the Reynolds-stress tensor has the following form:

$$\tau_{ij}^R = \mu_t \left( \frac{\partial u_i}{\partial x_j} + \frac{\partial u_j}{\partial x_i} - \frac{2}{3} \delta_{ij} \frac{\partial u_k}{\partial x_k} \right) \tag{9}$$

Here, $\delta_{ij}$ is the Kronecker delta function (it is equal to unity when $i=j$, and zero otherwise), μ is the dynamic viscosity, $\mu_t$ is the turbulent eddy viscosity coefficient and $k$ is the turbulent kinetic energy. In the frame of $k$-ε turbulence model, the turbulent eddy viscosity is given by

$$\mu_t = f_\mu \frac{C_\mu \rho k^2}{\varepsilon} \tag{10}$$

where $C_\mu$ is an empirical constant and ε is the turbulent dissipation. $f_\mu$ is turbulent viscosity factor which is defined as follows:

$$f_\mu = (1 - \exp(-0.0165R_y))^2 . (1 + \frac{20.5}{R_T}) \tag{11}$$

Where $R_T = \frac{\rho k^2}{\mu \varepsilon}$, $R_y = \frac{\rho \sqrt{k} y}{\mu}$ and $y$ is the distance from the wall. This function allows to take into account the laminar-turbulent transition. It is worth noting that $\mu_t$ and $k$ are zero for laminar flow. Also, if a computational cell is located away from a wall, $f_\mu$ will depend solely on $R_T$ .

Second-order spatial scheme is used for the faces that are common for two adjacent control volumes. For convective fluxes, the upwind approach is used. Time-implicit approximations of the continuity and momentum equations are used together with an operator-splitting technique. Due to difference of flow velocity and mesh size in the different regions of the computational domain, an adaptive time step algorithm with minimum time step of 0.001 second was used to ensure that the courant number ($\frac{u \Delta t}{\Delta x}$) is smaller than unity in all computational cells.

Mesh independence study was performed by monitoring the jet velocity at nozzle exit center for steady commanded piston speeds in the range *0.1 in/s* to *1.0 in/s* in increments of *0.1 in/s* similar to the experimental tests conducted by Krueger 2001. The ratio of jet velocity at the center of the nozzle exit to the average jet velocity across the nozzle exit ($\frac{u_{cl}}{U_j}$) denoted jet centerline velocity ratio are computed with different mesh sizes and then compared to the same variable obtained by PIV as reported in Krueger, 2001. Figure 4 shows the mentioned results for three mesh resolutions with the size of 231000, 400,000 and 3 million cells respectively in comparison with the PIV data.

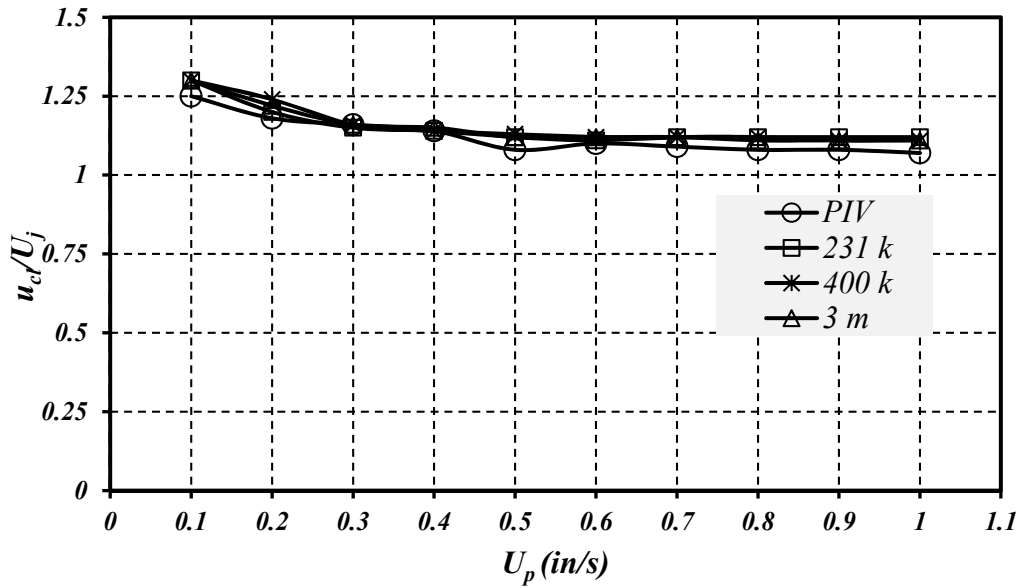

Figure 4. Comparison of computational and PIV measurements of $\frac{u_{cl}}{U_j}$ for different mesh resolutions at different piston speeds. The PIV data are taken from Krueger, 2001.

As can be seen, the jet centerline velocity ratio is closely following the one obtained from PIV measurements with the minimum and maximum difference of 2% and 4% respectively for the finest mesh resolution. The computational grid with 3 million cells provides the least difference to the PIV results for the jet centerline velocity ratio and therefore this mesh size is selected for all simulations. The spatial resolution in the selected mesh size is 0.5 mm in axial and radial directions. This is important since it helps to capture small counter-rotating vortices during the vortex ring generation. It should be noted that spatial resolution for PIV tests was about *1.23 mm* and *1.4 mm* in axial and radial directions respectively as reported in Krueger 2001. Also, the uncertainty for velocity results was reported as 1%.

It should be noted that the focus of this study is the relative performance of starting jets to its equivalent steady jets in terms of thrust and nozzle exit pressure contribution to the thrust. As was presented, much efforts have been spent to ensure the numerical accuracy of the results. However, if there is any uncertainty in the simulation results due to the employed numerical schemes or mesh resolution, it should be deemed acceptable in this study since those inaccuracies equally applied to the calculation of discussed parameters for both starting jets and their equivalent steady jets.

## 4. Results and discussions
### 4.1 Vortex ring formation
For numerical tests to generate starting jets, positive and negative slope triangular jet velocity programs, as illustrated in figure 2, are employed. Maximum jet velocity, $U_M$, is fixed at 1.03 m/s in all numerical tests. $\frac{t_m}{T}$ is chosen as 0.2 and 0.8 for NS and PS velocity programs respectively to replicate the experimental tests in Krueger 2001. It is worth noting that in the mentioned velocity programs, $T$ was increased from 0.05 seconds to 0.124 seconds to change $L/D$ from 2 to 5. Therefore, the jet velocity acceleration and deceleration are different for each $L/D$. Additionally, an impulsive jet velocity program with the constant piston speed resulting in the average jet velocity of $U_M/2$ is used at identical $L/D$ to study the effect of this type of jet velocity program on hydrodynamic properties of starting jets including thrust and nozzle exit over-pressure. Figure 5 shows the vorticity field of starting jets generated for $L/D=2-5$ under NS, PS and impulsive velocity programs at the end of jet termination. As can be seen, the general shape of generated vortex rings is somewhat similar for all velocity programs. However, the vortex rings are more elongated for NS program compared to those which were generated using PS and impulsive programs. As a result, the vortex rings are closer to nozzle exit for PS and impulsive programs at identical $L/Ds$.

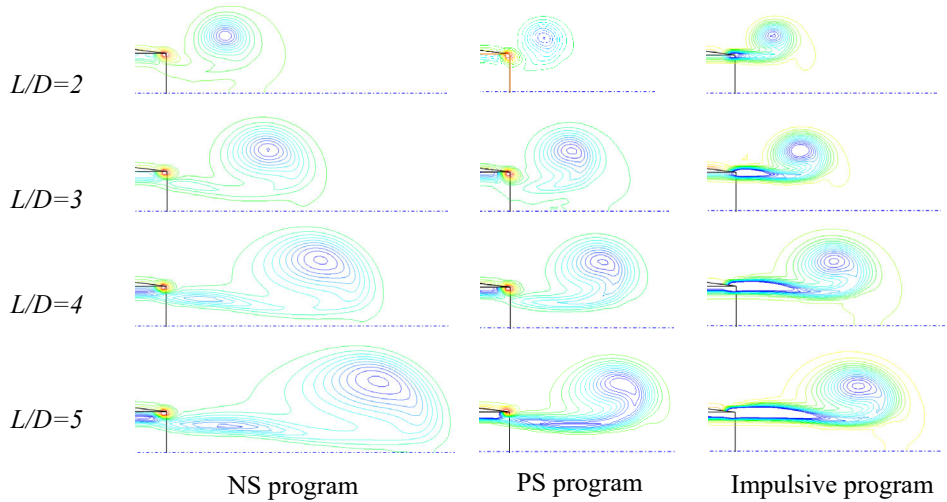

Figure 5. Vorticity field of starting jets generated by NS, PS and impulsive jet velocity programs for rang of *L/D*. The maximum axial length scale (*x/D*) for NS, PS and impulsive jet velocity programs are 4, 2 and 3 respectively.

For example, the vortex ring at *L/D=5* for PS and impulsive programs is about *1D* closer to the nozzle exit when compared with the vortex ring generated by NS program at the same *L/D*. In agreement with this finding, Krueger and Gharib 2003 reported a similar sequence shift of vortex rings in PS program relative to NS program. Also, a small secondary vortex with opposite vorticity near nozzle exit is seen in tested cases related to jet velocity programs. Figure 6 shows the time evolution of the main and secondary vortex for the starting jet at *L/D=2* generated by NS, PS, and impulsive velocity programs.

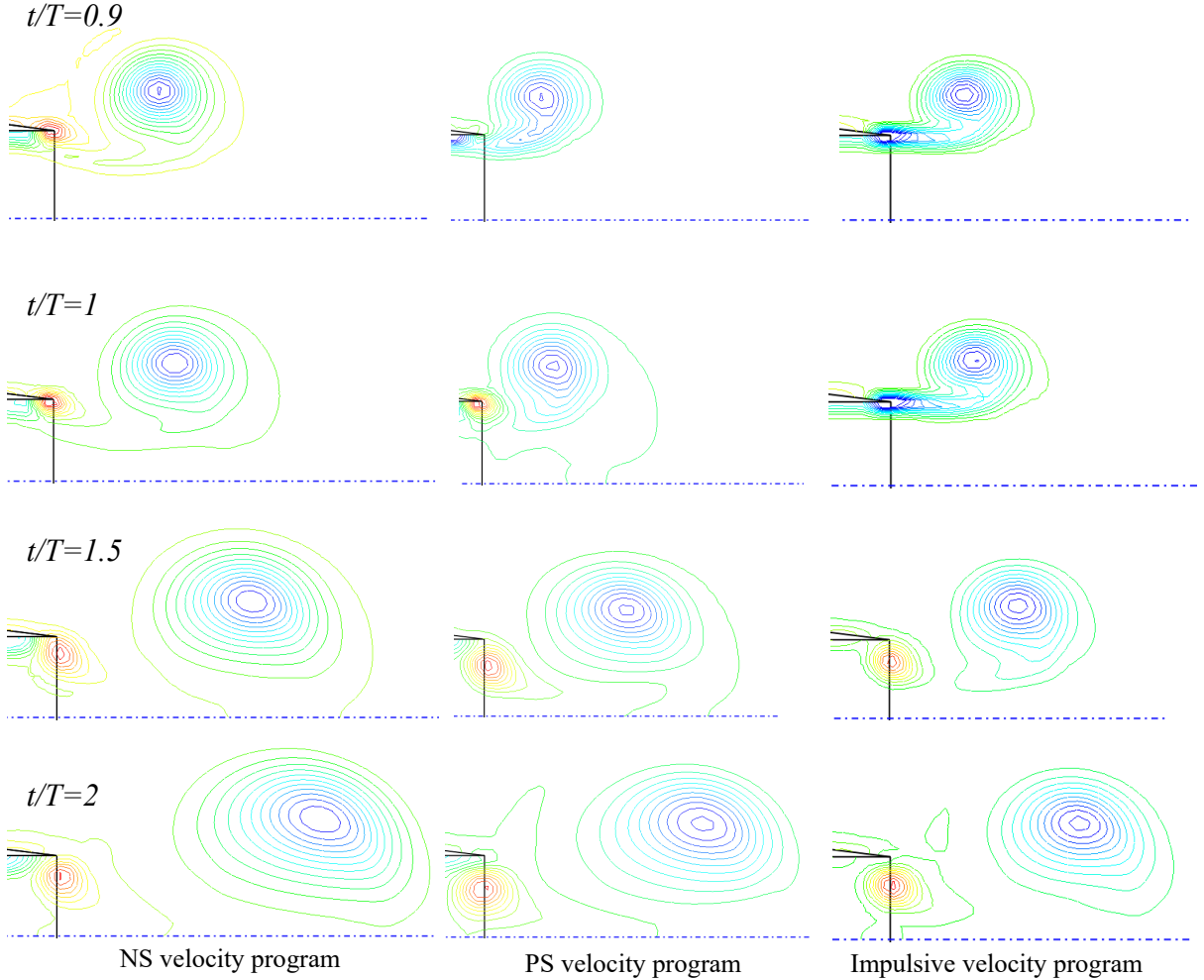

Figure 6. Vorticity contours related to the time evolution of the main vortex ring (mainly blue color) and secondary vortex (mainly red color) for the starting jet with *L/D=2* generated by NS, PS, and impulsive velocity programs. The vorticity direction of the main vortex ring and secondary vortex is counterclockwise and clockwise respectively. Maximum vorticity values of main and secondary vortices for the respective jet programs are different at each time step and can be found in the table (1). The maximum axial length scale (*x/D*) for NS, PS and impulsive jet velocity programs are 2.1, 1.875 and 1.7 respectively.

Table (1) lists the maximum vorticity values corresponding to the time steps presented in figure 6 for both the main vortex ring and secondary vortex. As can be seen, the earliest and latest secondary vortex starts to emerge at *t/T=0.9* and *t/T=1.5* for NS and impulsive programs respectively. Similar vortices are reported by Didden 1979 who generated experimental starting jets using a near impulsive velocity program. Also, recently, Fu & Liu 2015 numerically simulated the tests in Didden 1979 and observed the same vortices. The secondary vortex started to form after the jet termination in tests by Didden, 1979 in agreement with the presented results for the impulsive program. The secondary vortex grows as time progresses and reaches about $D/2$ in diameter at $t/T=2$ for all tested programs.

Table 1. Maximum vorticity values (in *1/s*) of main vortex ring and secondary vortex for NS, PS and impulsive velocity programs at different time steps

| $t/T$ | NS program | | PS program | | Impulsive program | |
|---|---|---|---|---|---|---|
| | Main | Secondary | Main | Secondary | Main | Secondary |
| *0.9* | *-370* | *140* | *-480* | *0* | *-325* | *0* |
| *1* | *-325* | *235* | *-430* | *520* | *-310* | *0* |
| *1.5* | *-190* | *112* | *-250* | *220* | *-220* | *195* |
| *2* | *-130* | *80* | *-170* | *160* | *-165* | *150* |

The numerical values presented in table 1 indicate that the vorticity magnitudes decrease for both main and secondary vortices as time advances. However, the secondary vortex in the PS program has the highest vorticity magnitude due to the higher deceleration in the respective jet velocity. It seems that there is a direct relationship between the secondary vortex and decelerating portion of the jet velocity program as it occurs during or after this time. It should be noted that although the piston speed is constant during jet ejection in the impulsive program, there is a jet velocity deceleration as piston stops at jet termination triggering the secondary vortex. Referring to equation (3), jet deceleration can cause under-pressure at nozzle exit which can be linked to the creation of the mentioned secondary vortex. In the next subsection, the results for computed nozzle exit pressure are presented revealing more quantitative information about this subject. Interestingly, the second counter-rotating vortex at the nozzle exit can be considered a manifestation of this under-pressure effect which is overlooked in previous studies including Krueger & Gharib 2003. Furthermore, considering that the total impulse can be calculated from the following equation (12), it indicates the reduction in jet impulse and thrust due to negative counter-rotating vortices reflecting the under-pressure effect.

$$I = \rho\pi \int \omega_\theta r^2 dr dx \qquad (12)$$

where $\omega_\theta$ is the azimuthal component of the vorticity field. Also, this relatively strong counter-rotating vortex causes negative jet velocity at nozzle exit centerline after jet termination. Figure 7 shows time-dependent normalized axial jet velocity recorded at nozzle exit center during and after termination of the starting jet at $L/D=2$ for all tested jet velocity programs. As can be seen, the jet velocity at nozzle exit follows closely the commanded piston velocity during jet ejection up to jet termination time. However, it continues to decline reaching negative values for all tested velocity programs. The jet velocity for PS and impulsive programs decreases with much higher deceleration reaching about 20% of maximum jet velocity ($U_M$) in magnitude near $t/T=2.5$. This significant negative jet velocity is due to the presence of strong counter-rotating secondary vortex at the nozzle exit which draws the flow inside the nozzle creating a negative flow velocity near the centerline. Negative jet velocities after jet termination were recorded for each tested $L/D$ in all of the velocity programs. However, the highest magnitude of negative jet velocity was only reached at $L/D=2$ for PS and impulsive velocity programs due to the higher vorticity strength of the respected secondary vortices. For the impulsive velocity program, similar negative velocities were observed in all range of tested $L/D$ since the jet deceleration magnitude was the same at jet termination.

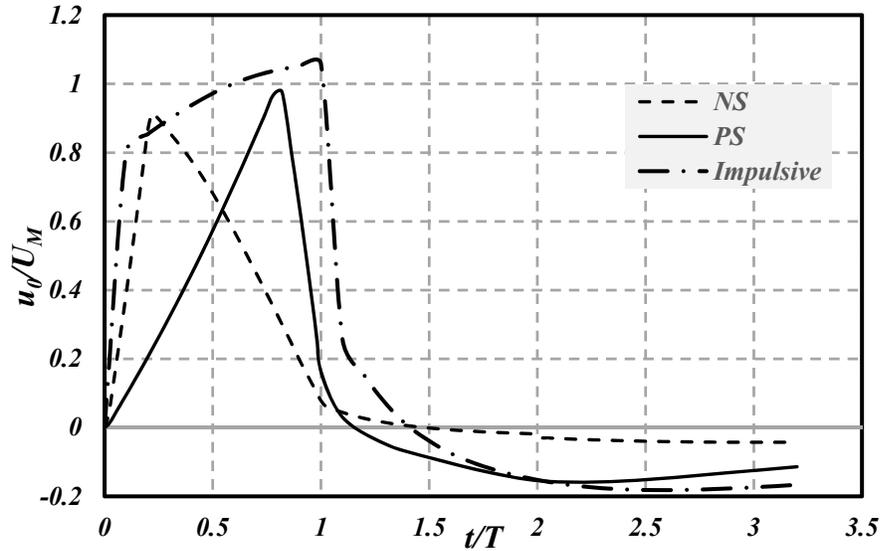

Figure 7. Normalized jet centerline velocity versus normalized time for NS, PS, and impulsive velocity programs at *L/D=2*

To visualize the effect of counter-rotating vortex in creating the discussed negative jet velocity after jet termination, figure 8 shows the velocity vector field demonstrating the flow reversal at nozzle exit for *L/D=2* at *t/T=2* for impulsive velocity program. The flow reversal at nozzle exit due to the secondary vortex may not be important in this study since it occurs after jet termination. But, as Krueger 2001 mentioned it may have caused some reduction in thrust and impulse of a fully pulsed jet at *L/D=2* where the vorticity field of preceding jet affects the newly formed vortex ring.

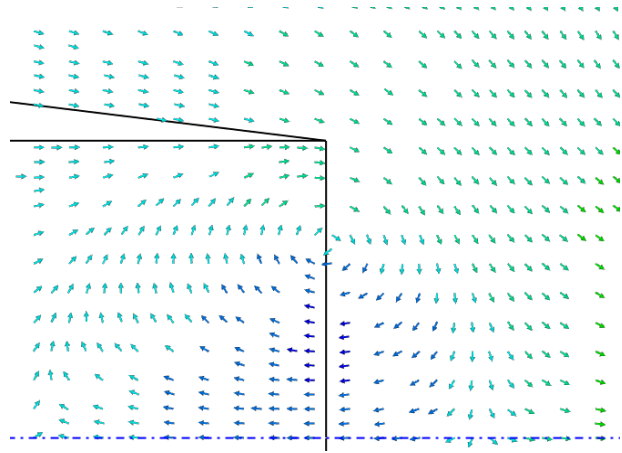

Figure 8. Velocity vector field of starting jet near nozzle exit generated by the impulsive program at *L/D=2* and *t/T=2*

### 4.2 Thrust ratio and nozzle exit pressure

Thrust ratio is calculated using equation (6). The time-averaged thrust for starting jets, $F_{sj}$, is obtained by summation of momentum and pressure thrusts denoted as $F_u$ and $F_p$ respectively which are computed by numerical integration of axial jet velocity and nozzle exit (gauge) pressure respectively over nozzle exit area for the duration of jet ejection time at each *L/D* for the respective jet velocity programs. The time-averaged thrust for the equivalent hypothetical steady je, $F_u$, is the same as momentum thrust which was obtained in previous step. Figure 9 presents the thrust ratio, $F_A$, for the range of studied $L/D$ and velocity programs. As can be seen, the thrust ratio for $L/D<4$ is lower than 1 for all studied velocity programs in contradiction to results presented in Krueger and Gharib 2003. Also, the thrust ration approaches 1 and slightly higher as $L/D$ increases noting it at $L/D=5$ for NS velocity program. From equation (6), it is clear that the trend of thrust ration is a function of competition between the pressure and momentum thrusts. To explain this trend, these parameters are investigated at each $L/D$ and respective velocity program.

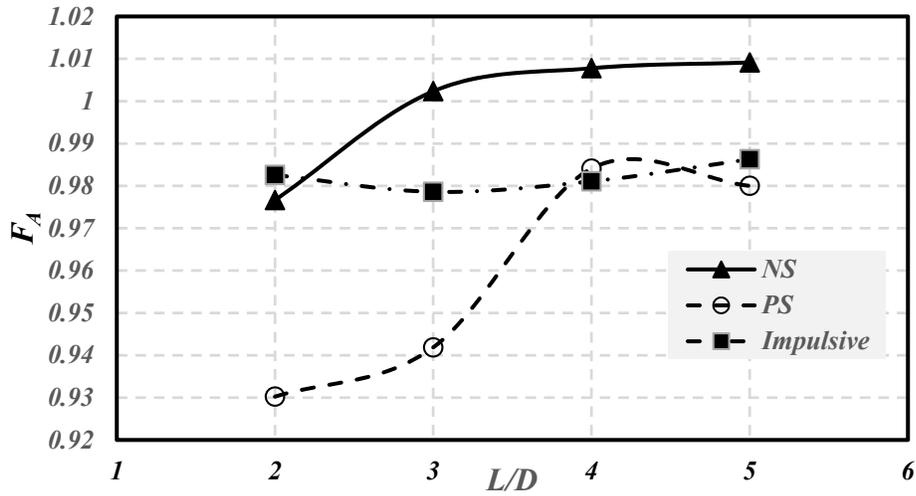

Figure 9. Thrust ratio for the range of $L/D$ at different velocity programs

It is found that both the momentum and pressure portions of thrust increases with $L/D$ but with different rates. For example, the momentum thrust increases from 42.8 *mN* to 43.75 *mN* (about 1 *mN* increase) as $L/D$ increases from 2 to 5 in NS velocity program. Nearly similar trends are observed in PS and impulsive velocity programs. Figure 10 shows comparison of the normalized axial jet velocity profile for $L/D=2$ and $L/D=5$ generated by the NS program at $t=t_m$ along with its normalized equivalent steady jet (with identical mass flux) for comparison. It is noted that the velocity profile features are similar to what was measured experimentally in Didden 1979 including an overshoot in the velocity near the wall. As can be seen, jet velocity magnitude near the jet core region ($r/R<0.8$) at $L/D=5$ is slightly higher than that at $L/D=2$. This pattern is seen during most of jet ejection time especially for $t>t_m$ making average momentum thrust slightly higher at $L/D=5$ compared to the one at $L/D=2$. This is mainly because the axial jet velocity profile at $L/D=5$ has relatively more time to develop into a nearly parabolic profile compared to that at $L/D=2$.

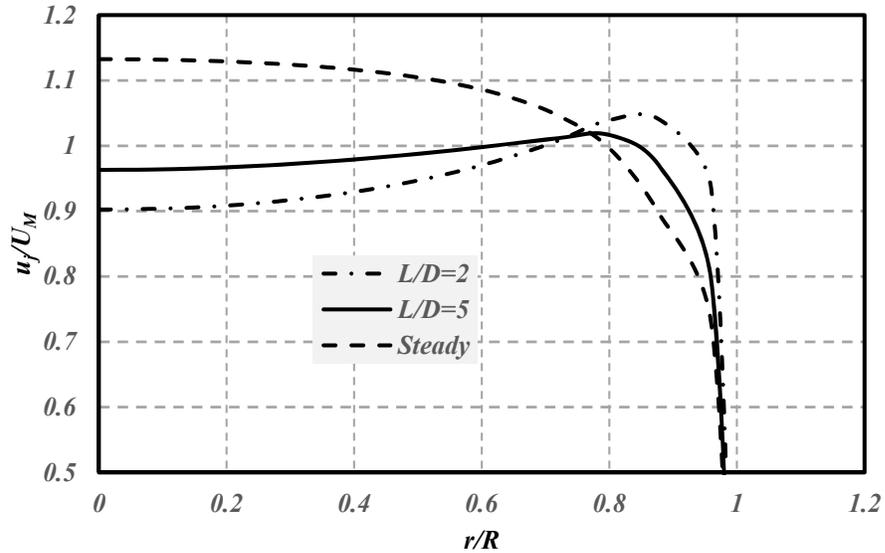

Figure 10. Comparison of normalized axial jet velocity at *L/D=2,5* generated by NS program along with its equivalent steady jet at *t=t_m*

To study the effect of pressure thrust component on the thrust augmentation ratio trend, the results for normalized average pressure thrust, $\frac{F_p}{F_u}$, are plotted in figure 11 for the range of *L/D* at different velocity programs. As can be seen, the normalized pressure thrust results start off with negative values at *L/D=2* for all velocity programs and increases with *L/D* for NS and PS programs with different rates. However, this trend is almost flat for the impulsive program. At higher *L/D*, the normalized pressure thrust reaches $\pm 0.01$ depending on the jet velocity program indicating its negligible contribution in the total thrust.

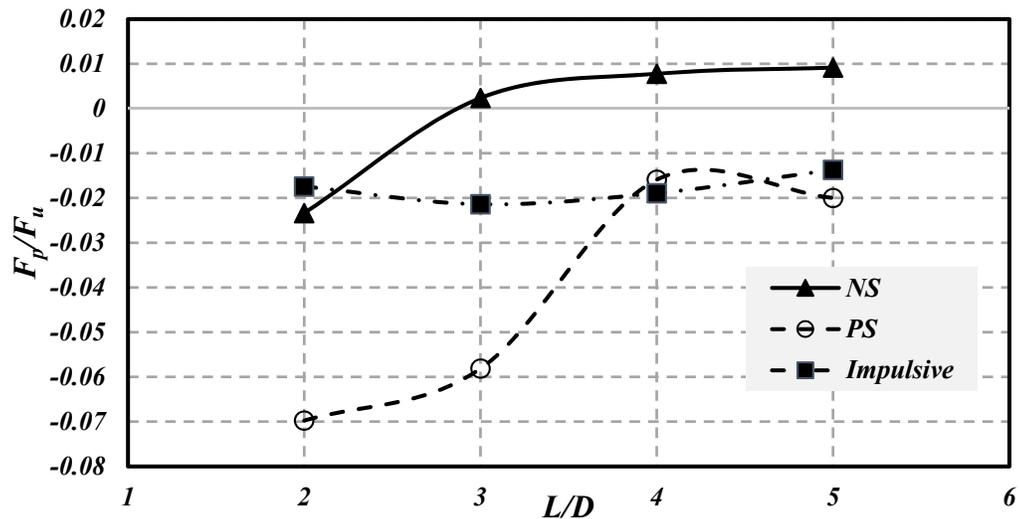

Figure 11. Normalized pressure thrust for the range of *L/D* at different jet velocity programs

Figure 11 shows that the normalized pressure thrust in the PS program has greater negative values compared to the one in the NS program for the studied range of *L/D*. To understand this trend, knowledge of nozzle exit pressure variation during jet ejection is necessary. Figure 12 presents the normalized pressure, $\frac{p_0-p_\infty}{\frac{1}{2}\rho U_M^2}$ , at nozzle exit center during normalized jet ejection time, $\frac{t}{T}$, for NS, PS, and impulsive jet velocity programs for the studied range of *L/D*. The pressure at nozzle exit center represents the behavior of pressure profile at nozzle exit during the jet ejection. It is worth noting that for the impulsive velocity program, the shape and magnitude of normalized nozzle exit pressure are the same for all studied *L/D* due to the similarity of jet velocity program during its initiation and termination as well as during jet ejection. Therefore, the normalized nozzle exit pressure is plotted only for one *L/D* to represent the rest of tested cases for this program. In agreement with the prediction made based on equation (3), the normalized nozzle exit pressure becomes positive or negative as jet velocity accelerates or decelerates respectively for NS and PS velocity programs. For impulsive velocity program, nozzle exit experiences a pressure maximum as the jet initiates and accelerates the flow from rest. However, it becomes negative for the rest of jet time as jet ejects the flow with constant piston speed. Figure 12 reveals that the magnitude of nozzle exit over-pressure portions in the NS velocity program is higher than that in the PS velocity program. The jet velocity acceleration $(\frac{\partial u}{\partial t})$ in NS velocity program is higher than that in the PS velocity program during acceleration portion of the jet and referring to equation (3), this results in a relatively higher nozzle exit over-pressure. Similarly, the magnitude of nozzle exit under-pressure portions in the NS velocity programs is lower than that in the PS velocity program due to lower jet velocity deceleration in NS velocity program compared to that in the PS program at the respected jet ejection time (*t>t_m*). The combination of these two effects makes the pressure thrust in NS velocity program slightly higher and less negative compared to PS velocity program (as shown in figure 11) resulting in a higher thrust ratio for the NS velocity program as seen in figure 9. For the impulsive velocity program, it seems that the nozzle under-pressure effect is more dominant due to the short time period of nozzle exit over-pressure. This results in small negative pressure thrust which does not vary considerably with *L/D* since the jet initiation and ejection velocities are identical for all *L/D* as stated before.

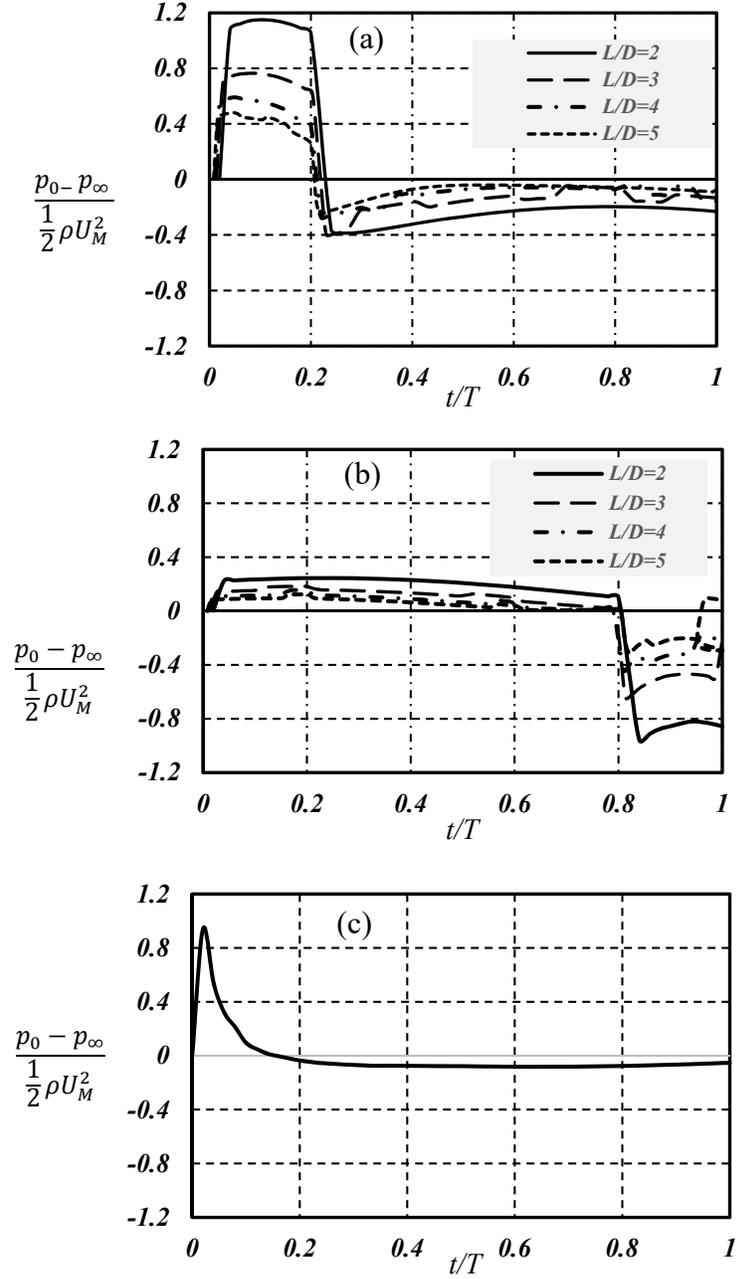

Figure 12. Normalized time-dependent nozzle exit pressure for (a) NS, (b) PS and (c) impulsive velocity programs for the range of *L/D*.

To validate the previous predictions based on equation (3), $\int_0^\infty \frac{\partial u}{\partial t} dx$ (the main unknown of this equation) is computationally calculated using the centerline jet velocity field for all velocity programs. Figure 13 shows this parameter versus normalized time for *L/D=2* (as a representative *L/D*) for all discussed velocity programs. The computational domain for calculation of $\int_0^\infty \frac{\partial u}{\partial t} dx$ is extended to length of *4D* from nozzle exit to capture enough centerline velocity data. As can be seen, $\int_0^\infty \frac{\partial u}{\partial t} dx$ is positive as the jet is in acceleration modes for both NS and PS velocity programs. Also, as jet starts decelerating, it decreases sharply but remains positive for a time duration which is different between NS and PS velocity program considering the different deceleration magnitudes. Then, it becomes negative during rest of jet deceleration for both PS and NS jet velocity programs. The presented numerical values confirm that $\int_0^\infty \frac{\partial u}{\partial t} dx > \frac{1}{2} u_0^2$ during jet acceleration resulting in nozzle exit over-pressure and $\int_0^\infty \frac{\partial u}{\partial t} dx < \frac{1}{2} u_0^2$ during jet deceleration resulting in nozzle exit under-pressure in agreement with presented results in figure 12. Also, the magnitude of $\int_0^\infty \frac{\partial u}{\partial t} dx$ is proportional to commanded jet velocity program acceleration or deceleration and determines the magnitude of nozzle exit pressure. For the impulsive velocity program, $\int_0^\infty \frac{\partial u}{\partial t} dx$ starts with a relatively large magnitude due to the initial high jet acceleration which results in the pressure spike as seen in Figure 12(c). Then, it remains slightly positive during the rest of jet ejection as piston pushes the fluid with constant speed. It is confirmed that $\int_0^\infty \frac{\partial u}{\partial t} dx < \frac{1}{2} u_0^2$ during the rest of jet ejection resulting in a slight nozzle exit under-pressure as shown in figure 12(c). It is worth noting that the nozzle over-pressure or under-pressure calculated from equation (3) is slightly different from those obtained from CFD since viscosity is ignored in deriving equation (3).

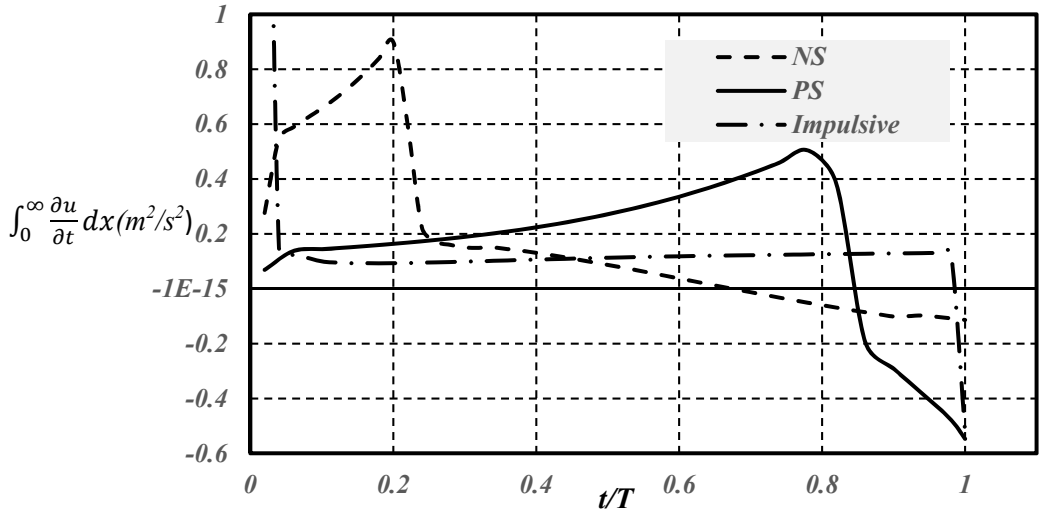

Figure 13. Variation of $\int_0^\infty \frac{\partial u}{\partial t} dx$ versus normalized time for $L/D{=}2$ at different jet velocity programs

Overall, it can be concluded from the presented results that the effect of nozzle exit pressure on the thrust of starting jets under the presented boundary conditions is slightly negative and at best is negligible. This is in contradiction with the conclusion presented in Krueger & Gharib 2003 regarding the effect of nozzle exit over-pressure on the thrust augmentation of starting jets. To find the reason for this dispute, the method that impulse and thrust were measured in their experimental tests was reviewed. It is found that the total impulse and average thrust of each single pulse were accomplished for each velocity program by calculating $I = \int_0^\infty F_{FB}(t)dt$ and $F_T = \frac{I}{t_p}$ respectively where $F_{FB}$ was the force measured by a force balance during starting jet ejection. The issue that can be seen in the provided equations is that the time interval in the calculation of impulse extends to infinity and not to the jet termination time ($t_p$). However, this is not the case for thrust calculation since it is obtained by dividing the calculated impulse by jet termination time. At first glance, this does not seem troubling since the thrust magnitude after jet termination should be close to zero, but it is possible that additional noise has been recorded during this extended time making the total average thrust artificially higher. As a result, Krueger & Gharib 2003 obtained that $I - I_U > 0$ ($I_U$ represents the velocity impulse) and concluded that the impulse surplus should come from pressure impulse but never measured or computed the nozzle exit pressure field to confirm this assumption. Also, it seems that the effect of counter-rotating vortex was not considered during the calculation of impulse using PIV data. Besides these issues, instrument and sensor errors can be another source of uncertainty, especially when dealing with very low magnitude forces in the scale of *mN*.

## 5. Summary and conclusions

Thrust and nozzle exit pressure in starting jets from a tube nozzle generated by different velocity programs at the range of *L/D=2-5* were numerically investigated. The thrust of starting jet was compared to that of steady jet based on identical jet velocity program at the nozzle exit without considering the flow unsteady effects.

The results for the thrust ratio showed little or no thrust augmentation for starting jets generated by different velocity programs in the range of studied *L/D*. Investigating the nozzle exit pressure revealed that nozzle experiences both over-pressure and under-pressure in accelerating and decelerating portions of jet velocity respectively in NS and PS jet velocity programs. This resulted in negative or negligible time-averaged pressure thrust which in turn reduced the thrust ratio specifically for starting jets generated by the PS program. Also, nozzle exit experienced a slight under-pressure during jet ejection in impulsive jet velocity programs. Moreover, secondary counter-rotating vortices at nozzle exit were observed during jet ejection and after jet termination in all studied velocity programs. The presence of the secondary vortex was linked to the nozzle exit under-pressure causing reduced impulse due to their opposite vorticity rotation relative to the main vortex ring. Furthermore, the secondary vortex caused flow reversal with jet velocity magnitudes as high as *0.2U_M* after jet termination at *L/D=2* in PS and impulsive velocity programs.

It should be noted that the generated thrust from starting or pulsed jets is still higher than that for steady jets with identical mass flow rate despite negligible contribution of nozzle exit pressure to the total thrust. This additional thrust is due to jet velocity program and its momentum and can be simply shown that the thrust augmentation ratio is 1.33 for triangular jet velocity programs disregarding their NS or PS ramps. Krueger and Gharib 2005 obtained similar ratio for NS velocity programs in the range of tested *L/D* from 2 to 6.

The Reynolds number based on the maximum jet velocity in this study is *1.3×10⁴* indicating the turbulent jet flow. Therefore, the presented case can represent the behavior of the nozzle exit pressure in the turbulent flow. For laminar starting jets, the vortex ring is more elongated in axial direction without a clear pinch-off compared to the one in turbulent flow as shown in Moslemi and Krueger, 2011. However, the same trend in terms of contribution of nozzle exit pressure to the thrust is expected since the discussed flow variables in equation (3) are independent of flow regime. It should be interesting to repeat this study with laminar starting jets to verify this prediction.

The presented nozzle configuration consists of a sharp cone angle of 7° (figure 3) to minimize the effect of generated vortex ring interaction with the nozzle following Krueger & Gharib 2003. This angle can influence the fluid entrainment by the jet and thereby affecting the centerline jet velocity portion which is induced by formation of vortex ring. This angle can increase to 90° making the nozzle blunted and the effect of its variation on the presented results can be an interesting subject for future studies.

The next logical question when studying thrust of starting jets in comparison to steady jets is that which propulsion scheme provides higher propulsive efficiency? That is whether the benefit of generating higher thrust by pulsed jet outweighs its energy cost? A few experimental studies including Moslemi and Krueger 2010 and Ruiz et al. 2011 reported higher propulsive efficiency for pulsed jet compared to steady jet at some test conditions. However, this topic is worth to be investigated again for future studies in the light of this paper results regarding the negligible effect of nozzle exit over-pressure contribution to the thrust of starting jets.

**Acknowledgements**
The author would like to sincerely thank Dr. Rouzbeh Riazi for editing the paper and providing valuable feedbacks about its content.